\documentclass[
twocolumn,
]{ceurart}
\pdfoutput=1
\usepackage{verbatim}
\usepackage{todonotes}
\usepackage{amsthm}
\newtheorem{definition}{Definition}

\usepackage{listings}
\begin{document}

\copyrightyear{2022}
\copyrightclause{Copyright for this paper by its authors.
  Use permitted under Creative Commons License Attribution 4.0
  International (CC BY 4.0).}

\conference{AIofAI 2022: 2nd Workshop on Adverse Impacts and Collateral Effects of Artificial Intelligence Technologies}

\title{Good AI for Good:\\How AI Strategies of the Nordic Countries Address the Sustainable Development Goals} 

\author[1]{Andreas Theodorou}[%
email=andreas.theodorou@umu.se
]
\cormark[1]
\address[1]{Department of Computing Science, Umeå University,Umeå  901 87 Sweden}

\author[1]{Juan Carlos Nieves}[%
email=juan.carlos.nieves@umu.se
]

\author[1]{Virginia Dignum}[%
email=virginia.dignum@umu.se
]

\cortext[1]{Corresponding author.}

\begin{abstract}
Developed and used responsibly Artificial Intelligence (AI) is a force for global sustainable development. Given this opportunity, we expect that the many of the existing guidelines and recommendations for trustworthy or responsible AI will provide explicit guidance on how AI can contribute to the achievement of United Nations' Sustainable Development Goals (SDGs). This would in particular be the case for the AI strategies of the Nordic countries, at least given their high ranking and overall political focus when it comes to the achievement of the SDGs. In this paper, we present an analysis of existing AI recommendations from 10 different countries or organisations based on topic modelling techniques to identify how much these strategy documents refer to the SDGs. The analysis shows no significant difference on how much these documents refer to SDGs. Moreover, the Nordic countries are not different from the others albeit their long-term commitment to SDGs. More importantly, references to \textit{gender equality} (SDG 5) and \textit{inequality} (SDG 10), as well as references to environmental impact of AI development and use, and in particular the consequences for life on earth, are notably missing from the guidelines.
\end{abstract}


\maketitle

\section{Introduction} \label{Introduction}

Since the launch of the United Nations' (UN) 2030 Agenda for Sustainable Development \cite{united_nations_transforming_2015}, many actions have been launched to analyse, realise, and discuss how this blueprint for peace and prosperity for people and the planet, can be achieved now and into the future.
The core of the UN's 2030 Agenda is the collection of 17 Sustainable Development Goals (SDGs), which describe concrete areas and steps for action to be taken by all countries. 
Across the whole line, while the world is making progress in some areas, we are also falling behind in delivering the SDGs overall \cite{the_sustainable_development_goals_report_sustainable_2021}. 

At the same time, Artificial Intelligence (AI) is increasingly becoming part of our daily lives. In previous work, we have contributed to a comprehensive analysis of the role of AI in achieving the Sustainable Development Goals \cite{vinuesa2020role}, concluding that it has the potential to shape the delivery of all 17 goals and 169 targets, both in a positive or a negative way. In fact, through a consensus-based expert elicitation process, Vinuesa {\em et al.} identified that ``AI can enable the accomplishment of 134 targets across all the goals, but it may also inhibit 59 targets''~\cite{vinuesa2020role}.

The contribution of AI towards the achievement of the SDGs is shaped by the vision and actions of national governments and international institutions. Research and development strategies for regulatory measures and oversight mechanisms for AI-based technologies affect directly how AI will contribute positively or negatively, or completely ignore an opportunity to enable sustainable development. Current strategies and proclamations are meant to ensure trust by creating long-term societal stability through the use of well-established tools and design practices, and as such can ensure well-being for all in a sustainable world.

The Nordic countries are often placed at the top of global rankings where it comes to the 2030 Agenda and the achievement of the SDGs \cite{sachs_decade_2021}. Moreover, like all countries in the Europe Union have released their own AI strategies; promising that socio-ethical values will guide any development and deployment of systems in their countries \cite{dexe_nordic_2020}. It is therefore relevant to see if AI strategies from the Nordic countries stress the potential of AI to support this effort, and how they compare to other countries. 

In this article, we put these strategies to the test. We developed a vocabulary of terms associated with SDG values. Then, by using common information retrieval techniques, we examine the relevance of those terms to our policies. Most importantly, we present not only our results but also our critical reflections of those results as we explore what exactly should be \textit{AI for good} by grounding the term into SDGs metrics and responsible AI practices. Even though this paper focuses on the Nordic countries, the text analysis method introduced can be to evaluate how other AI strategy documents address the SDGs. 

In the next section, we present the background and other related work to the study presented in this paper. Next, we discuss the methodology used and then present our results. In the penultimate section, we provide a critical discussion of our findings. We conclude this paper with a summary of our findings and planned future work. 
 
 \section{Background and Related Work}

In this section, we discuss the relevant background and similar work done presented in the literature.

\subsection{AI Governance}
The global movement around Responsible AI has become a main direction in AI research and practice. Countries, international organisations and business are all publishing a growing number of strategies and guidelines describing how to ensure that the development and use of AI meets ethical and societal principles \cite{theodorou_towards_2020}. 
Most recently, an 36 policy documents grouped socio-ethical values into 8 key themes: 1) \textit{privacy}; 2) \textit{accountability}; 3) \textit{safety and security;} 4) \textit{transparency and explainability}; 5) \textit{fairness and non-discrimination}; 6) \textit{human control of technology}; 7) \textit{professional responsibility}; and 8) \textit{promotion of human value} \cite{fjeld_principled_2020}. The authors searched the policy documents for the 47 keywords that make up the themes by using a purpose sampling method. In their analysis, Fjeld {\em et al.} noted how guidelines often call for increased access to technology and stakeholder involvement in its design. A more comprehensive analysis of 84 policy documents shows that there is a global convergence around five ethical principles: 1) \textit{Transparency}; 2) \textit{justice and fairness}; 3) \textit{non-Maleficence}; 4) \textit{responsibility}; and 5) \textit{privacy} \cite{jobin2019global}. 

Within the Nordic strategies, Robinson notes ``... the cultural values of trust, transparency, and openness have shaped discussions about AI.'' \cite{robinson_trust_2020}.
Across the Nordic countries, societies are based on the same fundamental values such as democracy, human rights, and sustainability. Over the years, these shared positions of strength have been used to establish synergies and to share experiences in a way that facilitates effective responses to a wide range of issues; providing benefits for all the people of the Nordic countries. With respect to AI development and use, a recent report by the Association of Nordic Engineers indicates that engineers have not only accepted their own responsibilities towards society but have also been pushing towards the minimisation of negative consequences of technologies \cite{podgaiska_nordic_2021}. They argue in favour of socio-technical solutions, where adequate governance can offer answers to specific questions such as ``Who benefits from the development of AI? Are those only a few individuals, specific groups, or a larger population?'' instead of generalised statements.

How exactly these principles are interpreted and applied still has a significant variance across the globe \cite{theodorou_towards_2020} and the academic literature; e.g. keywords such as \textit{transparency} and \textit{explainability} have multiple definitions\cite{arrieta_explainable_2019}. Even within the Nordics, i.e. countries with common cultural and historic background, the value of transparency differs policy to policy. For example, the Norwegian strategy, uniquely, considers access of citizens to government information as part of its goals for ensuring AI transparency, while Denmark and Finland focus more on data ethics, education (e.g. ``what is AI?''), and organisation aspects \cite{robinson_trust_2020}.

Unlike the studies presented above, our focus in this paper is not to explicitly check the occurrence of socio-ethical values. Rather, to examine the occurrence of keywords directly related to the SDGs and their targets as defined by the United Nations. However, an overlap of keywords, e.g. \textit{well-being}, is expected due to the socio-ethical nature of some of the SDGs.

\subsection{Sustainable Development Goals and AI}
In September 2015, the United Nations adopted the 2030 Agenda for Sustainable Development \cite{united_nations_transforming_2015}. Building on the principle of ``leaving no one behind”, the new Agenda emphasizes a holistic approach to achieving sustainable development for all. However, unlike AI strategies, the UN has produced 17 Sustainable Development Goals, listed in Table~\ref{tab:ontology}, each with its own concrete targets. The aim of the SDGs to both motivate and provide the means of assessing progress towards Sustainable Development.

The digitisation of our societies and overall technological progress can be seen as a catalyst for making a headway towards achieving sustainable development \cite{del_rio_castro_unleashing_2021}. Technological progress includes the development and usage of AI with showing that at least 59 of the sustainable development targets might actually be inhibited by \cite{vinuesa_role_2020}. Already, AI for Good efforts show the positive impact of AI on climate change, access to health and education, and inclusion. Accelerating these approaches requires principles and regulations to ensure that all AI is `good' AI: legal, ethical, robust, beneficial and verifiable. That, is while AI promises benefits for the SDGs, it will only be as positive and impactful as designed and utilised.

The potential for AI to help achieving UN's sustainable development goals is dependent on how responsible AI principles are addressed \cite{del_rio_castro_unleashing_2021}. Without solving problems of bias, energy consumption, power imbalances, and privacy associated with many current applications of AI, there is little hope that AI can be used as an instrument to achieve the SDGs \cite{astobiza_ai_2021}. Instead, we risk exemplifying negative narratives around AI, undermining trust and adoption of the technology\cite{sartori_sociotechnical_2022}, by even directly contributing negatively to SDGs.

\section{Methodology} \label{Methodology}
To present an evidence-based analysis of the SDGs and their presence in the AI strategies of the Nordic countries, the national AI strategies reports of the Nordic countries were mined.

\subsection{Topic mining model}\label{Subsection-TopicMining}

Our methodology follows a text mining approach \cite{aggarwal2012mining}. Text mining can be seen as a variation of data mining, that tries to find patterns from large databases. In text mining, the patterns are extracted from natural language text rather than from structured databases, image sets or other unstructured sets of data. In particular, we follow a topic modelling technique as a heuristic to analyse the vast amount of textual information that exits around AI strategies reports.  

Topic models are text mining models that provide a simple way to analyse large volumes of unlabelled text. Topics are usually defined by a finite set of words, which frequently occur in a text corpus. 

\begin{definition}
A topic model denoted by $M$ is defined as a tuple of the form $M = \langle t_1,\dots,t_m \rangle$ such that each $t_i (1 \leq i \leq n)$ is a finite set of words.
\end{definition}

One can understand each set of words of a topic model as probability distributions over the words in a vocabulary. Hence, one can build relationships between the vocabulary of a topic model and the vocabulary of a given set of documents. Naturally, the set of documents can be composed by a unique document. 

\begin{definition}
Let ${\cal L}$ be a finite space of words and $M = \langle t_1,\dots,t_m \rangle$ be a topic model. 
A mining model of $M$ w.r.t $\cal L$ is a tuple of the form  $\Lambda(M,{\cal L} ) =  \langle \sigma_1(t_1,{\cal L}),\dots,\sigma_m(t_m,{\cal L}) \rangle$ such that each $\sigma_i (t_i,{\cal L}) (1 \leq i \leq n)$ is a cumulative distribution function. 
\end{definition}

In the study presented in this paper, we have defined mainly two topic models of the Sustainable Development Goals.  One is defined by considering the keywords that appear in each name of the SDGs, and the second one is defined by considering synonyms of those words as identified in the Collins English Thesaurus and the keywords of the titles of the SDGs (see Table \ref{tab:ontology}). 

In the following section, we will describe different mining models that take as inputs the topic models of the SDGs (see Table \ref{tab:ontology}) and different finite spaces of words that were constructed by considering different national AI strategy reports. 

\subsection{Selection of Strategies}
There are over 600 AI-related policies released by prominent intergovernmental organisations, professional bodies, national-level committees and other public organisations, non-governmental, and private for-profit companies \cite{OECD}.  
We focused our efforts to the Nordic countries of Denmark \cite{ministry_of_finance_and_ministry_of_industry_national_2019}, Finland \cite{ministry_of_economic_affairs_and_employment_finlands_2017}, Norway \cite{ministry_of_local_government_and_modernisation_national_2020}, and Sweden \cite{ministry_of_enterprise_and_innovation_national_2018}\footnote{At time of our analysis, Iceland does not have a national AI strategy other than a webpage stating a declaration of intent to create such a strategy.}. In addition, we included the strategies from Germany \cite{federal_ministry_of_education_and_research_artificial_2018}, Japan \cite{japans_strategic_council_for_ai_technology_artificial_2017}, United Kingdom (UK) \cite{department_for_business_energy__industrial_strategy_industrial_2018}, and the United States of America (USA) \cite{national_science_and_technology_council_national_2016} to provide a comparison between the Nordics and the major `AI powerhouses' \cite{Deloitte}. We did not include China due to the lack of a national-level non-sectorial strategy available in English.

Finally, we complimented our corpus through the inclusion of the ``Guidelines for Trustworthy AI'' by the European Commission's High-Level Expert Group on AI \cite{european_commission_directorate-general_for_communications_networks_content_and_technolog_ethics_2019} and the IEEE ``Ethically-Aligned Design'' \cite{the_ieee_global_initiative_on_ethics_of_autonomous_and_intelligent_systems_ethically_2019}. The former serves as the basis of multiple strategies found in the European Union and the later is the product of the largest public consultation for an AI strategy and has been instrumental for the subsequent development of AI standards \cite{Winfield2018}.


\begin{table*}[t!]
    \centering
    \begin{tabular}{cp{0.3\linewidth}p{0.5\linewidth}}
        \toprule
        Goals as a topic & Topic model based on title keywords & Topic model based on the outputs of the Collins English Thesaurus and the keywords of the titles of SDGs\\
        \midrule
        G0 & Overarching terms & Sustainability, Sustainable Development Goal, SDG, Agenda 2030 \\
        G1 & Poverty& pennilessness, distress, necessity, hardship, insolvency, privation, penury, destitution, hand-to-mouth existence, beggary, indigence, pauperism, necessitousness\\
        G2& Hunger & hunger, undernutrition, maldutrition, starvation, famine, undernourishment, food \\ 
        G3 & Health, Well-being  & wellbeing, welfare, interest, health, benefit, advantage, comfort, happiness, prosperity\\ 
        G4 & Education & teaching, schooling, training, development, coaching, instruction, tutoring, tuition, indoctrination \\ 
        G5 & Gender & feminismm, sexism, women’s movement, suffragette, suffragist, feminist, sexist, emancipated \\
        G6 & Water, Sanitation & hygiene, cleanliness, sewerage, drinking water \\
        G7 & Clean Energy  &green energy \\ 
        G8 & Decent Work, Economic Growth &financial, business, trade, industrial, commercial, mercantile \\ 
        G9 & Industry, Innovation, Infrastructure & technological innovations \\ 
        G10 & Inequality & apartheid, linguistic imperialism, favouritism, bias, partiality, injustice, imbalance, nepotism \\
        G11 & Sustainable Cities, Sustainable Communities & Smart cities,  society, people, public, association, population, residents, commonwealth, general public, populace, body politic, state \\ 
        G12 & Responsible Consumption, Responsible Production & using up, waste, expenditure, exhaustion, depletion, utilization, dissipation, manufacture, manufacturing, construction, assembly, fabrication \\ 
        G14 & Life Below Water & biology, marine biology \\
        G15 &  Life on Land & agriculture \\
        G16 & Peace, Justice, Strong Institutions& truce, ceasefire, treaty, armistice, pacification, conciliation, cessation of hostilities,  fairness, equity, integrity, honesty, decency, impartiality, rectitude, reasonableness, uprightness, justness, rightfulness \\
        G17 & Partnerships, sustainable development & cooperation, association, alliance, sharing, union, connection, participation, copartnership \\ 
    \end{tabular}
    \caption{Topic models of the Sustainable Development Goals in terms of keywords. We have added `Goal 0' to cover overaching terms, e.g. \textit{SDG}, that may appear in policy documents. Synonyms for the words found in the title of the of the goals have been found by using the Collins' English Thesaurus\cite{Collins2015}. For the final search, both the titles and synonyms are used.}
    \label{tab:ontology}
\end{table*}

\subsection{Document Analysis}
For the purposes of analysing the selected corpus of AI guidelines, we applied the following steps:
\begin{itemize}
    \item Convert all PDF files into plain text to form our corpus.
    \item Stemming of the words
    \item Convert the resulting corpus into a vector of TF-IDF features.
    \item Search the TF-IDF features vector of each documents for our keywords.
    \item Sum TF-IDF scores for each SDG.
    \item Average the results and normalise.
    \item Construct a heatmap to visualse the results, by assigning colours based on the range of the results' values.
\end{itemize}

TF-IDF features extraction was handled by the \texttt{TfidfVectorizer} function offered by SciKit library \cite{scikit-learn}. The \texttt{TfidfVectorizer} was configured such as to apply a stop-words list, removing conjuctions and determiners, e.g. \textit{the}, \textit{and}, and common prepositions, e.g. \textit{towards}, from our text. The TfidfVectorizer was further modified to tokenise the documents using our own algorithm instead of the included one. By using the TextBlob \cite{loria2018textblob} and NLTK libraries \cite{Loper02nltk}, we are able to not only tokenise the text, but also stemmatise it. Moreover, our tokenisation algorithm ensured that we are counting not only single words, but also up to 2 \textit{n}-gram, i.e. contiguous sequences of two words (e.g. \textit{substainability goals}).
 

\section{Results}
\begin{figure*}[t!]
    \centering
    \includegraphics[width=\linewidth]{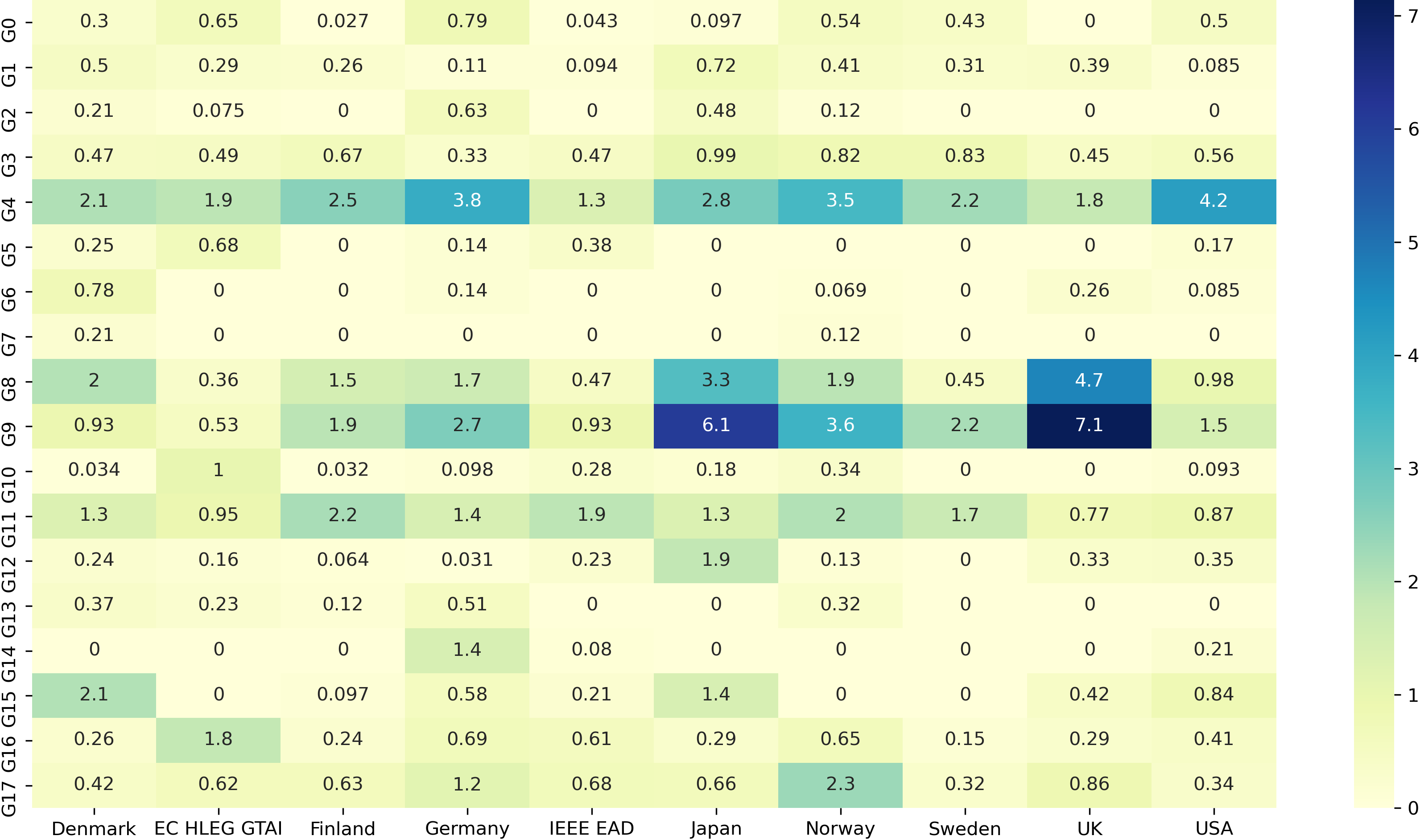}
    \caption{A heatmap of the TF-IDF values grouped for each goal
    of our keywords used to for each goal consult Table~\ref{tab:ontology}. Special attention needs to the be drawn to the fact that none of the policies analysed directly addresses Goal 5 and only Denmark's strategy makes explicit mentions to keywords associated with Goals 6 and 7.}
    \label{fig:heatmap}
\end{figure*}

From the 131 words and sequences of words used as keywords, only 74 were found in our corpus. Figure~\ref{fig:heatmap} shows the produced heatmap visualising the occurrence of the SDGs across the 10 policy documents selected. 

The Danish strategy is engaged with the most SDGs: 15 out of 17. They are followed by Germany and Japan that have references to 12 SDGs in their strategies. Finland, Norway, and the USA engage with 11 SDGs. Sweden and the UK mention just 9 SDGs. Both of our two international guidelines, the Guidelines for Trustworthy AI and IEEE Ethically-Alligned Design engage with 12 SDGs. Interestingly, only the UK makes no mention of our overaching terms \textit{Sustainable Development Goals} and \textit{SDG}.

Our analysis of the corpus, reveals an important gap in current guidelines and recommendations, that of addressing the impact of AI on natural resources, and specifically on SDGs 6 (water and sanitation), 7 (clean energy), 14 (life below water), and 15 (life on land). Furthermore, Goal 5 (gender) is completely ignored. On the contrary, Goals 4 (education), 8 (decent work and economic growth), and 11 (sustainable cities and communities) are the ones that most policy documents engaged with.

\section{Discussion}

  
At first glance, there is not much significant difference on how countries are doing in terms of references to SDGs. The Nordic countries did not differentiate themselves from others albeit their long-term commitment to SDGs and global ranking \cite{sachs_decade_2021}. 
 
Overall, the environmental impact of AI development and use---and in particular the consequences for life on earth (Goals 14 and 15)---are notably another aspect missing in the guidelines. As Kate Crawford exceptionally describes in her book `The Atlas of AI': AI is intrinsically connected with our natural resources \cite{crawford_atlas_2021}. Take as an example, the lithium and tin mines where the materials for semiconductors are mined at great human and environmental cost. In fact, natural resources, fuel, and human labour, are as core to AI systems as algorithms, data and infrastructures.

Denmark stands as an outlier from all other policies evaluated. It is the only country with \textit{some} direct references to Goals 6 (water and sanitation) and 7 (clean energy). These results are not a surprise given Denmark's commitment to those two goals as its top two priority areas for achieving 2030 Agenda \cite{ministry_of_foreign_affairs_handlingsplan_2017}. 
However, similarly, Goal 7 was also identified as a priority area by Finland \cite{annukka_path2030_2019}, but references to it are absent from their policy. Sweden also considers energy a challenging SDG, but also one which it is making significant progress towards the goal \cite{ministry_of_finance_sweden_2017}. Still, the little-to-none references to SDGs related to environmental impact is alarming. Particularly as we are developing larger and larger data-driven models that consume increasingly more electricity \cite{bender_dangers_2021, strubell_energy_2020}.

While \textit{fairness} is a value often found in AI policy documents \cite{jobin2019global}, explicit references to \textit{gender equality} (Goal 5) and \textit{inequality} (Goal 10) are---surprisingly---missing from the corpus used in our study. At first glance, this may highlight a limitation of our approach---and any other similar study: we can only search and find connections to goals or values based on synonym terms that \textit{we have thought of}. In our case, we developed our vocabulary of terms based on the UN's 2030 Agenda2030. However, this also highlights not only the level of abstraction used in the writing of AI guidelines and recommendations, but also disconnect from other global policy efforts. Such texts mostly refer to high-level values, such as fairness without further exemplifying concrete examples of the concepts. As others have discussed, such high-level description stands potentially in the way of a shared interpretation of how to operationalise the guidelines \cite{theodorou_towards_2020,AlerTubella2019IJCAI}. 

 
These are much-needed efforts, but still much work is needed to ensure that all AI is developed and used in responsible ways that contribute to trust and well-being. Nevertheless, even though organisations agree on the need to consider ethical, legal and societal principles, how these are interpreted and applied in practice, varies significantly across the different recommendation documents. Arguably, due to the cultural nature of ethical values we will never agree on universal definitions and requirements \cite{turiel_culture_2001}. Yet, the SDGs provide at least some universal metrics meant to enable the long-term sustainability of our societies. Linking them with the values found in AI ethics policies could help us produce actionable and verifiable requirements for AI systems \cite{umbrello_mapping_2021}. Otherwise, we end up with so called `ethics washing,' where organisations can claim adherence to abstract values without ensuring that AI systems that contribute to social good \cite{bietti_ethics_2019}. 

Truby goes as far as to state ``Countries signing up to the SDGs could also refuse to adopt tech that does not abide by SDG principles. Countries have sufficient justification to do this, in that allowing non-compliant AI would negate progress toward the SDGs they have committed to achieving'' \cite{truby_governing_2020}. Going a step further, we argue that even that if we are unable to ensure the net-positive impact of a system to our society, then the most sustainable choice is to refrain from its development and deployment. After all, we are not obliged to use AI, but when do, we must do so for social good.

\section{Conclusions and Future work}
Our results are worrisome; there is a clear disconnect between AI strategies and Agenda 2030. While the socio-ethical values AI strategies often advocate are subject to cultural interpretations, a link to SDGs provide much needed concretisation of those values. Otherwise, we end up with `ethics washing' instead of actively ensuring that any AI developed, deployed, and used is done so for the social good.

From the methodology point of view, we argue that this study is the first step towards a deeper mining of the AI strategies reports. Our topic models lack a context analysis, i.e. our topic models do not observe in which way the SDGs are mentioned in the AI strategies reports. To extend our mining of the AI strategies reports, we aim to apply methods of argument mining in our future work. Argument mining offers features to recognize context and consistency in natural-language text information \cite{lawrence2020argument}. Finally, we will like to extend our study to include additional countries, e.g. all top 20 performing countries with regards to Agenda 2030.

\begin{acknowledgments}
This work was partially supported by the Wallenberg AI, Autonomous Systems and Software Program (WASP), funded by the Knut and Alice Wallenberg Foundation and by the European Commission’s Horizon2020 project HumanE-AI-Net (grant 952026).
\end{acknowledgments}

\bibliography{bib/ath_old,bib/bib}

\end{document}